\documentclass[prd,showpacs,superscriptaddress,preprintnumbers,nofootinbib]{revtex4}
\usepackage[latin9]{inputenc}
\usepackage{color}
\usepackage{amsmath}
\usepackage{amssymb}
\usepackage{graphicx}
\usepackage[unicode=true,
 bookmarks=false,
 breaklinks=false,pdfborder={0 0 1},backref=section,colorlinks=true]
 {hyperref}
\hypersetup{
 linkcolor=blue,urlcolor=blue,citecolor=red}

\makeatletter
\@ifundefined{textcolor}{}
{%
 \definecolor{BLACK}{gray}{0}
 \definecolor{WHITE}{gray}{1}
 \definecolor{RED}{rgb}{1,0,0}
 \definecolor{GREEN}{rgb}{0,1,0}
 \definecolor{BLUE}{rgb}{0,0,1}
 \definecolor{CYAN}{cmyk}{1,0,0,0}
 \definecolor{MAGENTA}{cmyk}{0,1,0,0}
 \definecolor{YELLOW}{cmyk}{0,0,1,0}
}

\makeatother

\begin{document}
\title{Two dimensional non-Hermitian harmonic oscillator: coherent states}
\author{Masoumeh Izadparast}
\email{masoumeh.izadparast@emu.edu.tr}

\author{S. Habib Mazharimousavi}
\email{habib.mazhari@emu.edu.tr}

\affiliation{Department of Physics, Faculty of Arts and Sciences, Eastern Mediterranean
University, Famagusta, North Cyprus via Mersin 10, Turkey}
\begin{abstract}
In this study, we introduce a two dimensional complex harmonic oscillator
potential with space and time reflection symmetries. The corresponding
time independent Schrödinger equation yields real eigenvalues with
complex eigenfunctions. We also construct the coherent state of the
system by using a superposition of 12 eigenfunctions. Using the complex
correspondence principle for the probability density we investigate
the possible modifications in the probability densities due to the
non-Hermitian aspect of the Hamiltonian. 
\end{abstract}
\date{\today }

\maketitle

\section{Introduction}

In quantum mechanics the Hamiltonian operator, $\hat{H}=\frac{\hat{P}^{2}}{2m}+V\left(\hat{x}\right),$
represents the energy of a quantum system. Imposing the energy of
the physical system to be real, for a long time, made Hermiticity
a necessary property for the Hamiltonian operator $\hat{H}$. In 1998,
Bender and Boettcher introduced the concept of non-Hermitian Hamiltonian
with parity and time symmetries which admits real energy spectra \cite{Benderrealspectra1998}.
This achievement has brought a new domain of physical complex operators
into the quantum mechanics which extends its boundaries both theoretically
\cite{Theoretical} and experimentally \cite{Experimental}.

\emph{Non-Hermitian Hamiltonian:}

Fundamental cornerstones in quantum theory are build on i) the reality
of the physical quantities such as energy, ii) the conservation of
the probability density. The latter implies the unitary of time evolution
of a quantum system. A quantum mechanical model is preserved as long
as these principles are satisfied. Here in non-Hermitian version of
quantum mechanics, the Hermiticity is replaced by a non-Hermitian
Hamiltonian with the similar property i.e., preserving the reality
of the physical quantities including the energy spectrum. Mathematically
speaking, a Hamiltonian may be non-Hermitian due to the complex form
of the potential function, i.e., $V(x)\neq V^{\ast}(x)$. This provides
a new approach in a wide diverse class of complex Hamiltonians. Denoting
the $\mathcal{P}$arity and $\mathcal{T}$ime symmetry, they appear
to be the best substitution for Hermiticity. Parity is the space reflector
operator with linearity property changes the sign of the coordinate
and momentum operators $\hat{x}$ and $\hat{p}$, respectively. But,
the anti-linear time operator merely reverses the sign of\ the imaginary
part. In short one writes 
\begin{equation}
\mathcal{P}\hat{x}\mathcal{P}=-\hat{x},\text{ }\mathcal{P}\hat{p}\mathcal{P}=-\hat{p},
\end{equation}
and

\begin{equation}
\mathcal{T}i\mathcal{T}=-i,\text{ }\mathcal{T}\hat{p}\mathcal{T}=-\hat{p}
\end{equation}
in which $\mathcal{P}$ and $\mathcal{T}$ are the Parity and Time
symmetry operators, respectively. In the non-Hermitian quantum theory
it is necessary for the Hamiltonian to be \textit{invariant} under
the $\mathcal{P}$arity and $\mathcal{T}$ime transformation which
is called $\mathcal{PT}$-symmetry. Although, $\mathcal{PT}$ and
Hamiltonian commute, one cannot expect that they have simultaneous
eigenfunctions because they are not Hermitian. Since, the $\mathcal{PT}$
is not linear, if the energy spectrum is real, then the $\mathcal{PT}$-symmetry
of the Hamiltonian remains unbroken \cite{bender2007making}. On the
other hand, for those $\mathcal{PT}$-symmetric Hamiltonians whose
energy eigenvalues are complex, the $\mathcal{PT}$-symmetry is broken.
The $\mathcal{PT}$-symmetric Hamiltonian is defined as 
\begin{equation}
\mathcal{PT}\hat{H}\left(\mathcal{PT}\right)^{-1}=\mathcal{PT}\hat{H}\mathcal{TP}=\hat{H}.
\end{equation}
As we mentioned before, there exists one more condition for any feasible
quantum theory to be satisfied which states that the norm of the wave
functions must be positive and invariant in time \cite{ZnojilConsevation2001}.
In non-Hermitian quantum theory, however, the associated probability
density may attain complex values. With this contradiction, three
conditions for the relevant probability density in the complex plane
for a particle with harmonic potential were imposed \cite{bender2010complex,bender2002complex}:
i) the infinitesimal measurement of the imaginary part of the probability
should be zero i.e., $\Im\left(\rho\left(z\right)dz\right)=0$, ii)
the real part of the probability must be positive i.e., $\Re\left(\rho\left(z\right)dz\right)\geq0$,
and finally iii) its integral over the whole space must be one i.e.,
$\int_{C}\rho\left(z\right)dz=1$.

Historically, Bender and Boettcher \cite{Benderrealspectra1998} proposed
a new approach to employ non-Hermitian Hamiltonians to find real spectra
which is the $\mathcal{PT}$-symmetric Hamiltonian. They introduced
a family of complex potentials (also \cite{fernandez1999family})
and clarified that the solutions of the Schrödinger equation are transferred
from the real to complex domain. In different study, they identified
and generalized the concept of $\mathcal{PT}$-symmetry in other senses.
Variety of potentials and mathematical efforts are done in \cite{bender1998,bender2005,bendercp2002,bender2002complex,bender2003,bender2007making,bender2003must}.
Furthermore, Mostafazadeh explained how the Hermiticity is replaced
with $\mathcal{PT}$-symmetry in \cite{Mostafazadeh2003exact}, by
introducing the concept of Pseudo-Hermiticity as a $\mathcal{PT}$-symmetric
Hamiltonian with discrete energy and complete biorthonormal eigenbasis
vectors. Besides, Mostafazadeh expanded the idea of pseudo-Hermitian
and its properties in \cite{Mostafazadeh:2001nr,Mostafazadeh2010pseudo,Mostafazadeh:2002id}.
Znojil in \cite{ZnojilGeneralization2001}, and relevantly in \cite{Znojillinearity2002}
studied real energy spectra for $\mathcal{PT}$-symmetric Hamiltonians
and their generalizations as nonlinear Hamiltonians. He investigated
a complexified harmonic oscillator \cite{znojil1999pt} and reviewed
solvability of various Hamiltonians with $\mathcal{PT}$-symmetry
in \cite{ZnojilSolv2002} and \cite{Znojilnewtyp2004}. Moreover,
he expressed pseudo-norms conservation in $\mathcal{PT}$-symmetry
with the spontaneous breaking symmetry and the examination of the
Coulomb and harmonic oscillator correlations in \cite{znojil2000coulomb}.
In this respect, a complex Lie algebra have been introduced to examine
non-Hermitian systems in \cite{BAGCHI2000285}, generalization and
modification of the continuity equation and the normalization of $\mathcal{PT}$-symmetric
quantum mechanics in \cite{Bagchi2001}, replacing $\mathcal{PT}$-asymmetry
with $\mathcal{CPT}$-symmetry in \cite{CalicetiEmanuela2005} and
a search for $\mathcal{PT}$-invariant potentials leading to real
spectra in \cite{Ahmed:2001na,levai2000systematic}. Furthermore,
in \cite{mazharimousavi2008non}, it is demonstrated that in the polar
coordinate system a $\mathcal{PT}$-symmetry can be consider as a
combination of a Hermitian and a non-Hermitian ($\mathcal{PT}$-symmetric)
Hamiltonian. The energy spectra of complexified Morse, Scarf-II, and
Pöschl-Teller potentials are discussed in \cite{MeyurSanjib}. In
the case of 2D harmonic oscillator, rationalizing method is employed
to demonstrate the 2D complex harmonic oscillator in the extended
phase space in \cite{VirdiJasvinder}.

\emph{Coherent states:}

Coherent state is mostly a subject of interest in quantum optics and
is used to express the superposition of a certain number of states
in a bounded quantum mechanical system. In fact, the average of energy
of several eigenstates in a quantum mechanical system represents the
energy of the correspondence classical model. Besides, the expectation
value of position and momentum of coherent states in quantum approach
demonstrate the classical behavior of the particle. The coherent state
for a harmonic oscillator was denoted first by E. Schrödinger in quantum
mechanics where he was investigating the correspondence principle.
Schrödinger described a coherent state as the summation of several
states underlying the annihilation operator exertion. In Dirac notation,
for a real Harmonic oscillator a coherent state may be shown as $\left\vert \beta\right\rangle =|\beta|e^{i\theta}$
where $|\beta|$ denotes the amplitude and $\theta$ is the phase
of $\left\vert \beta\right\rangle $ such that
\begin{equation}
\hat{a}\left\vert \beta\right\rangle =\beta\left\vert \beta\right\rangle ,\left\langle \beta|\beta\right\rangle =1
\end{equation}
in which $\hat{a}$ is the annihilation non-Hermitian operator and
$\beta$ is its complex eigenvalue. 
\begin{figure}
\includegraphics[scale=0.25]{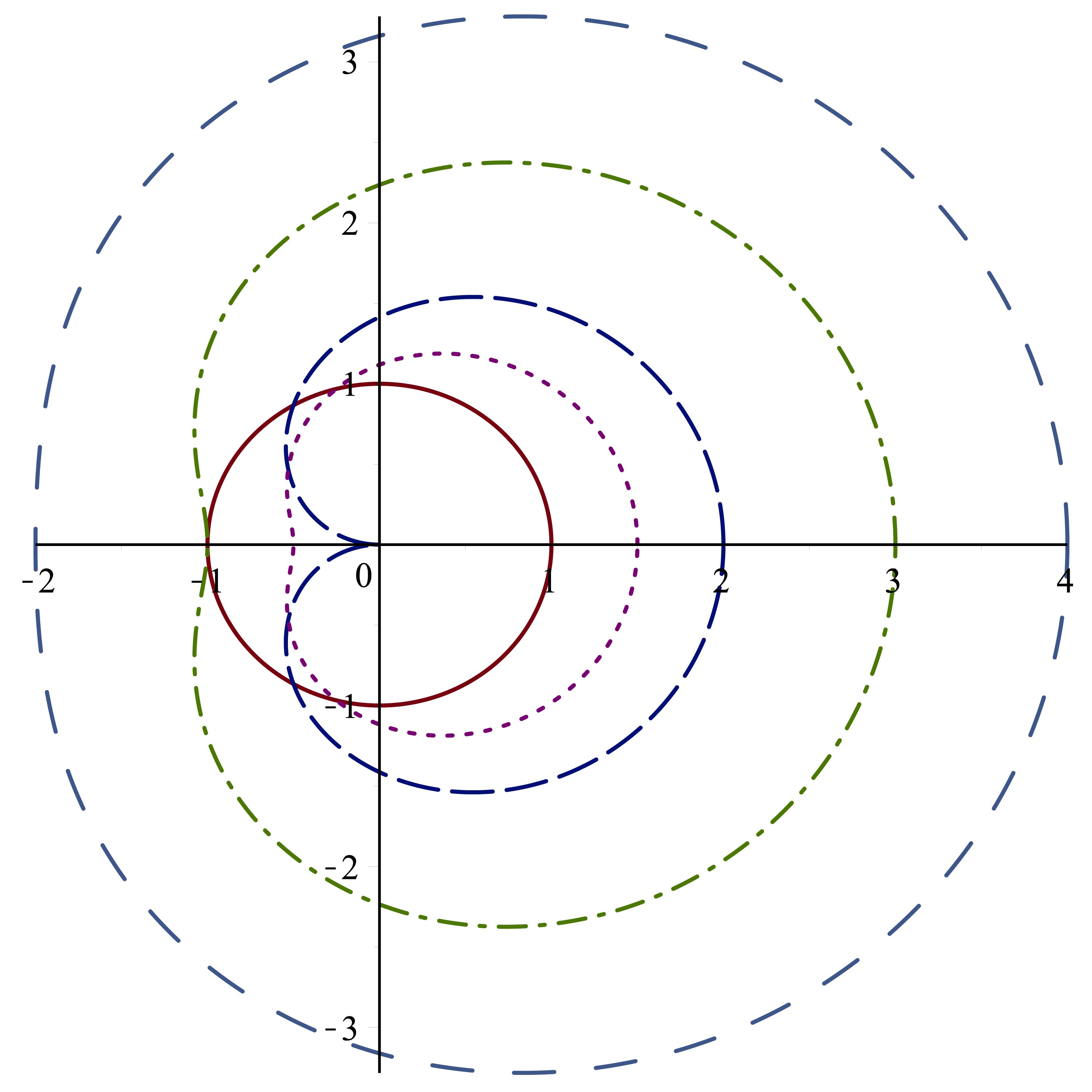} \caption{The behavior of the deformed harmonic oscillator potential (its absolute
value over $\dfrac{m\omega^{2}}{2}$) in terms of the polar angle
$\phi$ when $r=1$ and $\Lambda=0,0.5,1,2$ and $3$ corresponding
to SOLID,DOT,DASH,DASH-DOT and DASH-SPACE respectively. At the two
extremal limits where $\Lambda=0,\infty$ the potential becomes $\phi-$symmetric
but for the values of $\Lambda$ in between the potential depends
on $\phi$ critically. }
\end{figure}

In terms of the energy eigenkets of the harmonic oscillator (say $\left\vert n\right\rangle $)
, the representation of the coherent state is found to be 
\begin{equation}
\left\vert \beta\right\rangle =e^{-\frac{|\beta|^{2}}{2}}\sum_{n=0}^{\infty}\dfrac{\hat{a}^{n}}{\sqrt{n!}}\left\vert n\right\rangle .
\end{equation}
We note that, two different coherent states are not orthogonal, i.e.,
$\left\langle \beta|\beta^{\prime}\right\rangle \neq0.$ Glauber,
who was awarded for the Nobel prize in 2006 and his colleagues, in
\cite{GlauberRoy130.2529,GlauberRoy131.2766,louisell1973quantum,GlauberBook},
expressed the coherent states as the classical analogy of the radiation
in quantum optics. Gerry and Knight in \cite{Gerry1997} represented
the properties of the so-called Schrödinger-cat states and clarified
the field states in the electrodynamics aspect of the quantum optics.
The minimum uncertainty in time evolutionary form is presented in
\cite{howard1985minimum} by Howard and Roy who introduced the coherent
state of a harmonic oscillator \cite{HowardRoy}. In \cite{H. G. Oh},
the minimum uncertainty and likeliness of classically equation of
motion corresponding to the coherent states of a damped harmonic oscillator
is examined. The superposition of the harmonic oscillator considering
two different position dependent mass models has been investigated
in \cite{Biswas2009}. The classical aspect of the harmonic oscillator
is used to analogize coherent states of 2D harmonic oscillator in
vortex structure, in \cite{chen2003vortex}. Furthermore, let's mention
that the coherent states of the $PT$-symmetric quantum systems have
been studied in \cite{PTCS}.

In this present work, we introduce a $2$-dimensional complexified
harmonic oscillator resulting in real eigenvalues. In order to find
probability density, eigenfunctions are transferred from real plane
to complex plane. We examine the integral of probability density in
the space which ensures unity for the real and zero for imaginary
parts. Furthermore, we study superposition of 12 states to find corresponding
coherent state of the obtained wavefunctions.

Finally to complete the introduction, we would like to add that, most
of the lower dimensional quantum problems are considered as toy models
which shed light on the more complicated problems in the real three
dimensional quantum systems. Nevertheless, there are systems in three
dimensions which effectively can be reduced to two dimensions. For
a 2D harmonic oscillator we refer to the work of Li and Sebastian
\cite{Li2018} where the Landau quantum theory of a charged particle
in a uniform magnetic field has been considered. In their work, with
a specific magnetic vector potential, the problem is reduced to a
2D isotropic harmonic oscillator.

\section{Schrödinger equation and the 2D-complexified Harmonic oscillator}

We start with the two dimensional time-independent Schrödinger equation
with a presumed complex potential in the polar coordinate system given
by

\begin{equation}
-\dfrac{\hbar^{2}}{2m}\bigtriangledown^{2}\psi\left(r,\phi\right)+\dfrac{m\omega^{2}}{2}\left(\Lambda\dfrac{e^{i\phi}}{r}+1\right)r^{2}\psi\left(r,\phi\right)=E\psi\left(r,\phi\right),
\end{equation}
in which $m$ and $\omega$ are the mass and the angular frequency
of the harmonic oscillator and $\Lambda$ is a positive constant parameter.\\
In Fig. 1 we plot the absolute value of the deformed harmonic oscillator
potential over $\dfrac{m\omega^{2}}{2}$ in terms of $\phi$ for different
values of $\Lambda$ and $r=1.$ It is observed that a nonzero $\Lambda,$
particularly $0<\Lambda<1,$ modifies the behavior of the potential
(in terms of $\phi$) significantly.

To simplify and make the differential equation separable, we transfer
(6) from the polar to the Cartesian coordinates system where the potential
becomes
\begin{equation}
V\left(x,y\right)=\dfrac{m\omega^{2}}{2}\left(\Lambda\left(x+iy\right)+x^{2}+y^{2}\right).
\end{equation}
Based on the assumption of the $\mathcal{PT}$-symmetric Hamiltonian
/ potential, $V\left(x,y\right)$ shouldn't vary under the $\mathcal{PT}$-transformation.
Thus, the condition $V^{\ast}\left(-x,-y\right)=V\left(x,y\right)$
has to be hold. Apparently in (7) the $x$ component of the proposed
harmonic oscillator is not invariant under the parity reflection while
the $y$ segment completely supports the time and space symmetries.
This implies that $V^{\ast}\left(-x,-y\right)\neq V\left(x,y\right)$
and hence the potential is not $\mathcal{PT}$-symmetric. To cope
with this inconsistency, one can decompose the potential into $V_{x}\left(x\right)+V_{y}\left(y\right)$
as two combined one-dimensional oscillators. This yields a Hermitian
and a $\mathcal{PT}$-symmetric potential given by
\begin{equation}
V_{x}\left(x\right)=x^{2}+\Lambda x
\end{equation}
and
\begin{equation}
V_{y}\left(y\right)=y^{2}+i\Lambda y
\end{equation}
respectively. A similar treatment has been used in \cite{mazharimousavi2008non}
where the overall potential was called $\Pi\mathcal{T}$-symmetry
(instead of $\mathcal{PT}$-symmetry) such that $\Pi\mathcal{T}$-operator
is an invertible, non-Hermitian operator, consists of a time and phase
reflectors in polar coordinates, given by 
\begin{equation}
\Pi:\phi\rightarrow2\pi-\phi,\mathcal{T}:i\rightarrow-i.
\end{equation}
Introducing $\alpha=\frac{m\omega}{\hbar}$, after applying the separating
method on (6), one finds
\begin{equation}
-X^{^{\prime\prime}}+\alpha^{2}\left(\dfrac{\Lambda}{2}+x\right)^{2}X=k_{x}^{2}X
\end{equation}
and
\begin{equation}
-Y^{^{\prime\prime}}+\alpha^{2}\left(i\dfrac{\Lambda}{2}+y\right)^{2}Y=k_{y}^{2}Y.
\end{equation}
in which $\psi\left(x,y\right)=X\left(x\right)Y\left(y\right)$, $k_{x}^{2}+k_{y}^{2}=k^{2}$,
$k_{x}^{2}=\frac{2mE_{x}}{\hbar^{2}}$, $k_{y}^{2}=\frac{2mE_{y}}{\hbar^{2}}$
and $E_{x}+E_{y}=E.$ Furthermore, we apply a change of variables
expressed by $\Tilde{x}=\left(x+\frac{\Lambda}{2}\right)\sqrt{\alpha}$,
$\Tilde{y}=\left(y+i\frac{\Lambda}{2}\right)\sqrt{\alpha}$ and $\tilde{k}_{x,y}^{2}=k_{x,y}^{2}/\sqrt{\alpha}$
to simplify the above equations as
\begin{equation}
-X^{^{\prime\prime}}+\Tilde{x}^{2}X=\Tilde{k}_{x}^{2}X
\end{equation}
and
\begin{equation}
-Y^{^{\prime\prime}}+\Tilde{y}^{2}Y=\Tilde{k}_{y}^{2}Y.
\end{equation}
Now, we are dealing with two simple harmonic oscillators whose eigenvalues
and eigenvectors are known. Referring to any standard text-book in
quantum mechanics, one writes the full eigenvalues and eigenfunctions
of each coordinate as given by
\begin{equation}
X_{n}=C_{n}H_{n}e^{-\Tilde{x}^{2}/2}
\end{equation}
and
\begin{equation}
Y_{m}=C_{m}H_{m}e^{-\Tilde{y}^{2}/2},
\end{equation}
with their correspondence eigenvalues
\begin{equation}
E_{x,n}=\frac{\hbar\omega}{2}\left(2n+1\right)
\end{equation}
and
\begin{equation}
E_{y,m}=\frac{\hbar\omega}{2}\left(2m+1\right)
\end{equation}
in which $n,m=0,1,2,...$ and $H_{n}$/$H_{m}$ are the Hermite polynomials
while the constants $C_{n}$/$C_{m}$ are the normalization constants.

The energy of the real and complex part of the system behave as the
2D harmonic oscillator with the real potential \cite{chen2003vortex}.
The total eigenfunction is found to be
\begin{equation}
\psi_{nm}\left(\Tilde{x},\Tilde{y}\right)=\dfrac{1}{\sqrt{C_{nm}}}H_{n}\left(\Tilde{x}\right)H_{m}\left(\Tilde{y}\right)e^{-\left(\Tilde{y}^{2}+\Tilde{x}^{2}\right)/2}
\end{equation}
in which $\dfrac{1}{\sqrt{C_{nm}}}=C_{n}C_{m}.$ Since, the $\Pi\mathcal{T}$-symmetry
is preserved by the Hamiltonian (i.e., $[H,\Pi\mathcal{T}]=0$), Eq.
(19) is simultaneous eigenstates of the Hamiltonian and $\Pi\mathcal{T}$-operators.
To normalize the eigenfunctions, we refer to the redefinition of the
norm in the Hilbert space due to the implication of the non-Hermitian
Hamiltonian \cite{bender2010complex,bender2007making}. The inner
product of the two different eigenfunctions is defined as
\begin{equation}
\left\langle \psi_{nm}|\psi_{n^{\prime}m^{\prime}}\right\rangle =\int{dxdy\left(\Pi\mathcal{T}\psi_{nm}\right)}\psi_{n^{\prime}m^{\prime}}=\delta_{nn^{\prime}}\delta_{mm^{\prime}}\left(-1\right)^{m}.
\end{equation}
\\
Applying $\Pi\mathcal{T}$-operator on $\psi_{nm}$, $X_{n}$ and
$Y_{m}$ demonstrates different attribute. Based on the real argument,
$X_{n}$ does not vary under the $\Pi\mathcal{T}$ transformation.
Thus, the normalization follows the similar discussion extended in
\cite{Arfken} where it is shown that $\int_{-\infty}^{\infty}H_{n}^{2}e^{-\Tilde{x}^{2}}d\Tilde{x}=2^{n}n!\sqrt{\pi}$.
In the case of $Y_{m}$, the effect of $\Pi\mathcal{T}$-operator
on $H_{m}\left(y+i\frac{\Lambda}{2}\right)$ gives $H_{m}\left(-y-i\frac{\Lambda}{2}\right)=\left(-1\right)^{m}H_{m}\left(y+i\frac{\Lambda}{2}\right)$.
Therefore, $\left\langle \psi_{nm}|\psi_{nm}\right\rangle $ for odd
and even $m$ is negative and positive, respectively. What is the
physical interpretation of the negative norm in the Hilbert space?
It is identified as the Charge, Parity and Time symmetry ($\mathcal{C}\Pi\mathcal{T}$,
hereafter) to hold the symmetry of the Hamiltonian unbroken \cite{bender2002complex}.
The so-called charge operator $\left(\mathcal{C}\right)$ is introduced
to modify any theory with unbroken $\Pi\mathcal{T}$-symmetry. The
$\mathcal{C}$-operator is linear and represented in coordinate space
as a summation of simultaneous eigenfunctions of the Hamiltonian and
$\Pi\mathcal{T}$-operator \cite{bender2007making}. In fact, $\mathcal{C}$-operator
reverses the negative sign of odd $m$ in Eq. (20) to fulfill the
positivity of the norm in Hilbert space. The $\mathcal{P}$ and $\mathcal{C}$
operators do not equate because the parity is a real operator but
the charge is complex. Indeed, the Hermiticity convention in a quantum
mechanical model reforms to $\mathcal{C}\Pi\mathcal{T}$, where, 
\begin{equation}
\left(\mathcal{C}\Pi\mathcal{T}\right)\hat{H}\left(\mathcal{C}\Pi\mathcal{T}\right)=\hat{H}
\end{equation}
and therefore
\begin{equation}
\left\langle \psi_{nm}|\psi_{nm}\right\rangle _{\mathcal{C}\Pi\mathcal{T}}=\int_{-\infty}^{\infty}{dx}\int_{C}{dy\left(\mathcal{C}\Pi\mathcal{T}\psi_{nm}\right)\psi_{nm}}.
\end{equation}
Herein $C$ is the integration contour for the complex $y$ coordinate
which is the real $y$ axis shifted down on the imaginary axis as
of $y\rightarrow y-i\frac{\Lambda}{2}.$ Consequently the $\mathcal{C}\Pi\mathcal{T}$-norm
becomes
\begin{equation}
\left\langle \psi_{nm}|\psi_{nm}\right\rangle _{\mathcal{C}\Pi\mathcal{T}}=\int_{-\infty}^{\infty}{dx\int_{-\infty-i\frac{\Lambda}{2}}^{\infty-i\frac{\Lambda}{2}}dy\left(\mathcal{C}\Pi\mathcal{T}\psi_{nm}\left(x,y\right)\right)}\psi_{nm}\left(x,y\right).
\end{equation}
Furthermore, as ${\mathcal{C}\Pi\mathcal{T}\psi_{nm}\left(x,y\right)=}\psi_{nm}\left(x,y\right)$
the change of variable $y=\tilde{y}-i\frac{\Lambda}{2}$ yields
\begin{equation}
\left\langle \psi_{nm}|\psi_{nm}\right\rangle _{\mathcal{C}\Pi\mathcal{T}}=\int_{-\infty}^{\infty}{dx\int_{-\infty}^{\infty}d\tilde{y}\left(\psi_{nm}\left(x,\tilde{y}\right)\right)}^{2}
\end{equation}
which upon the fact that $\psi_{nm}\left(x,\tilde{y}\right)$ is a
real function it yields
\begin{equation}
\left\langle \psi_{nm}|\psi_{nm}\right\rangle _{\mathcal{C}\Pi\mathcal{T}}=\int_{-\infty}^{\infty}{\int_{-\infty}^{\infty}dxd\tilde{y}\left\vert \psi_{nm}\left(x,\tilde{y}\right)\right\vert }^{2}=1.
\end{equation}
This is because the latter equation is the standard normalization
relation for the 2D harmonic oscillator. As of the side result, the
normalization relation (23) suggests that the probability density
to be defined as
\begin{equation}
\rho\left(x,y\right)={\left(\mathcal{C}\Pi\mathcal{T}\psi_{nm}\left(x,y\right)\right)}\psi_{nm}\left(x,y\right).
\end{equation}
This $\rho\left(x,y\right)$ with the real $x$ and complex $y$ satisfies
the complex correspondence principle for the probability density on
the specific contour $C$ \cite{bender2010complex,bender2002complex}.

\section{The coherent states of the 2D-Harmonic oscillator}

In a quantum system, relation between classical and quantum mechanical
viewpoints has been one of the substantial topics to study. The coherent
state is distinguished as an superposition of numerous quantum mechanical
states, having minimized uncertainty with the mean energy of the correspondence
states which are not orthogonal. Here we resemble a classical 2D harmonic
oscillatory motion with the corresponding quantum mechanical circumstances.
Applying variation method on the classical Lagrangian, gives the equations
of motion of a particle undergoing the 2D complexified harmonic oscillator
potential (7) with solutions given by 
\begin{equation}
\Tilde{x}=|\beta|\sqrt{\frac{\hbar}{2m\omega}}\cos\left(\omega t-\theta_{x}\right)
\end{equation}
and 
\begin{equation}
\Tilde{y}=|\gamma|\sqrt{\frac{\hbar}{2m\omega}}\cos\left(\omega t-\theta_{y}\right).
\end{equation}
The classical position of the particle given in Eq.s (27) and (28)
imply the oscillational behavior as described in the classical mechanical
domain. In fact, these equations yield an elliptical motion depending
on the amplitude and the phase difference of the correspondence components.
For the real potential, it represents a real elliptical motion \cite{chen2003vortex}
but as the imaginary term is added into this system, it deforms the
elliptical attribute. Furthermore, the coherent states of the 2D-harmonic
oscillator may be formed by a simple multiplication of the two 1D
corresponding coherent states, namely 
\begin{equation}
\Tilde{\Psi}_{\beta\gamma}\left(\Tilde{x},\Tilde{y}\right)=\left\langle \tilde{x}|\beta\right\rangle \left\langle \tilde{y}|\gamma\right\rangle .
\end{equation}
Herein, $\left\vert \beta\right\rangle $ and $\left\vert \gamma\right\rangle $
are the coherent states corresponding to $\tilde{x}$ and $\tilde{y}$
coordinates mentioned in equation (5), respectively. In the case of
a temporal coherent state, time in exponential regime is allocated
i.e., 
\begin{equation}
\Tilde{\Psi}_{\beta\gamma}\left(\Tilde{x},\Tilde{y},t\right)=e^{-\frac{1}{2}\left(|\beta|^{2}+|\gamma|^{2}\right)}\sum_{m=0}^{\infty}\sum_{n=0}^{\infty}\dfrac{\beta^{n}\gamma^{m}}{\sqrt{n!m!}}\psi_{nm}\left(\Tilde{x},\Tilde{y}\right)e^{-\hbar\omega\left(n+m+1\right)t}.
\end{equation}
Using Cauchy product for the two partial series, the time evolutionary
form of the 2D harmonic oscillator's coherent state, i.e., $\Tilde{\Psi}_{\beta\gamma}\left(\Tilde{x},\Tilde{y},t\right)$
reads as 
\begin{equation}
\Tilde{\Psi}_{\beta\gamma}\left(\Tilde{x},\Tilde{y},t\right)=e^{-\frac{1}{2}\left(|\beta|^{2}+|\gamma|^{2}\right)}\sum_{N=0}^{\infty}\sum_{K=0}^{N}\dfrac{\beta^{K}\gamma^{N-K}e^{-\hbar\omega\left(N+1\right)t}}{\sqrt{K!\left(N-K\right)!}}\psi_{K,N-K}\left(\Tilde{x},\Tilde{y}\right).
\end{equation}
In this respect $n=K,m=N-K$, and $\frac{\beta}{\gamma}=Ae^{i\theta}$
in which $A$ and $\theta$ represent the relative amplitude and phase
difference between $\tilde{x}$ and $\tilde{y}$ coordinates, respectively.
Finally we find the coherent state up to any $N$ given by
\begin{equation}
\Tilde{\Psi}_{\beta\gamma}\left(\Tilde{x},\Tilde{y},t\right)=\sum_{N=0}^{\infty}C_{N}\Phi_{N}\left(\Tilde{x},\Tilde{y}\right)e^{-\hbar\omega\left(N+1\right)t},
\end{equation}
where
\begin{equation}
\Phi_{N}\left(\Tilde{x},\Tilde{y}\right)=\left(\dfrac{1}{\sqrt{1+|A|^{2}}}\right)^{N}\sum_{K=0}^{N}\binom{N}{K}^{1/2}\left(Ae^{i\theta}\right)^{K}\psi_{K,N-K}\left(\Tilde{x},\Tilde{y}\right).
\end{equation}
Let's comment that, $\Phi_{N}\left(\Tilde{x},\Tilde{y}\right)$ expresses
the elliptical stationary coherent state of an oscillator whose phase,
amplitude, and also the constant $\Lambda$ give its final form \cite{chen2003vortex}.

\emph{Normalization:}

The normalization procedure mimics the $\mathcal{C}\Pi\mathcal{T}$-symmetry
which converts the odd terms signs. Ultimately, the time-independent
part of the coherent state is normalized to unity. In the other words,
application of $\mathcal{C}\Pi\mathcal{T}$ on the coherent state
is found to be 
\begin{equation}
\mathcal{C}\Pi\mathcal{T}\Phi_{N}=\left(\dfrac{1}{\sqrt{1+|A|^{2}}}\right)^{N}\sum_{K=0}^{N}\binom{N}{K}^{1/2}\left(Ae^{-i\theta}\right)^{K}\mathcal{C}\Pi\mathcal{T}\psi_{K,N-K}\left(x,y\right),
\end{equation}
which results in
\begin{equation}
\int_{-\infty}^{\infty}{dx}\int_{C}{dy}\left(\mathcal{C}\Pi\mathcal{T}\Phi_{N}\right)\Phi_{N}dxdy=1.
\end{equation}
\begin{figure}
\includegraphics[scale=0.25]{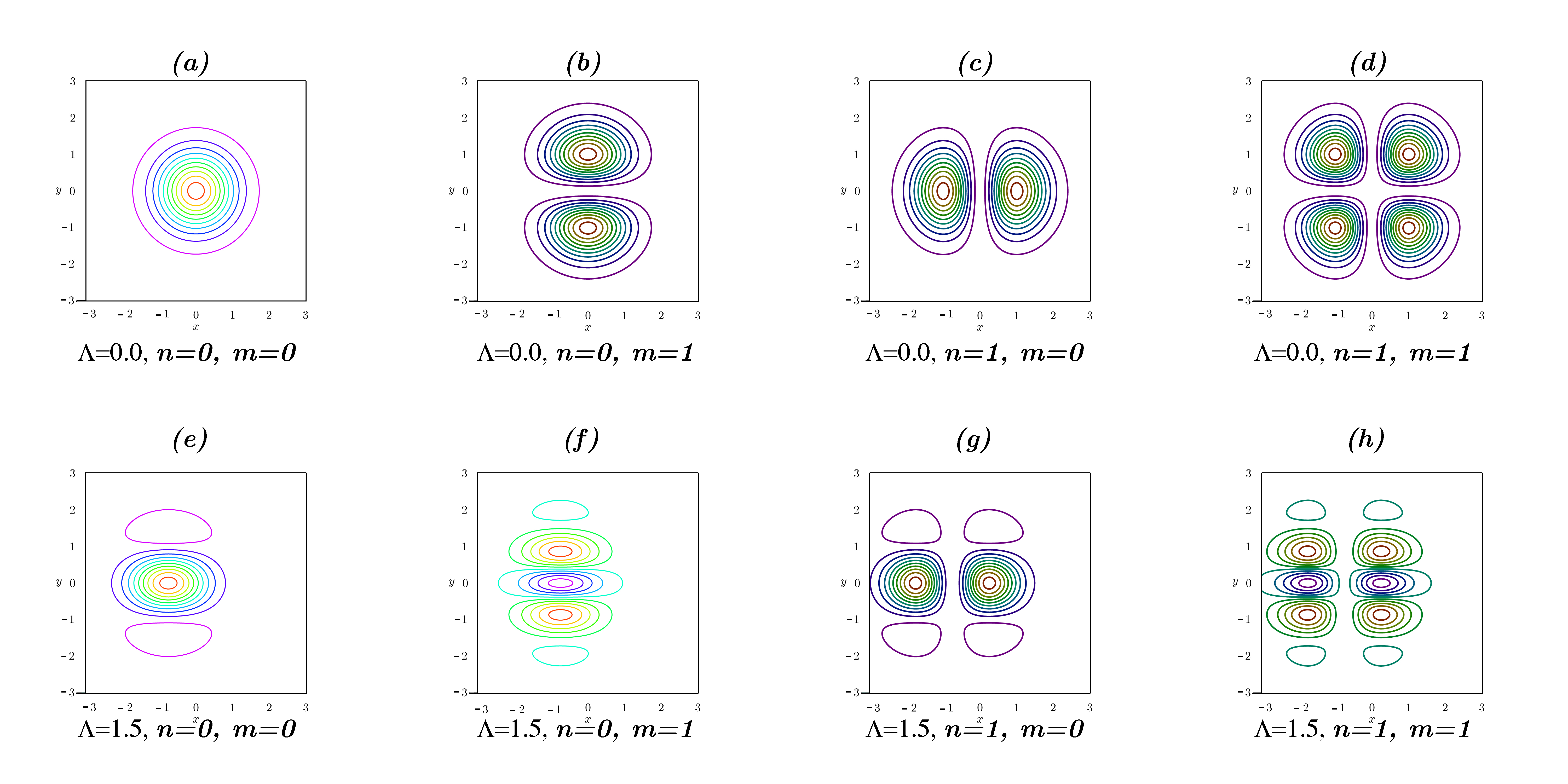} \caption{Plots of $\left\vert \psi_{00}\right\vert ^{2}$, $\left\vert \psi_{01}\right\vert ^{2}$,
$\left\vert \psi_{10}\right\vert ^{2}$and $\left\vert \psi_{11}\right\vert ^{2}$
with $\Lambda=0,$ from left to right respectively in the first row.
Plots of $\Re\left[\left(\psi_{00}\right)^{2}\right]$, $\Re\left[\left(\psi_{01}\right)^{2}\right]$,
$\Re\left[\left(\psi_{10}\right)^{2}\right]$ and $\Re\left[\left(\psi_{11}\right)^{2}\right]$
with $\Lambda=1.5$ from left to right respectively in the second
row. }
\end{figure}

\begin{figure}
\center \includegraphics[scale=0.25]{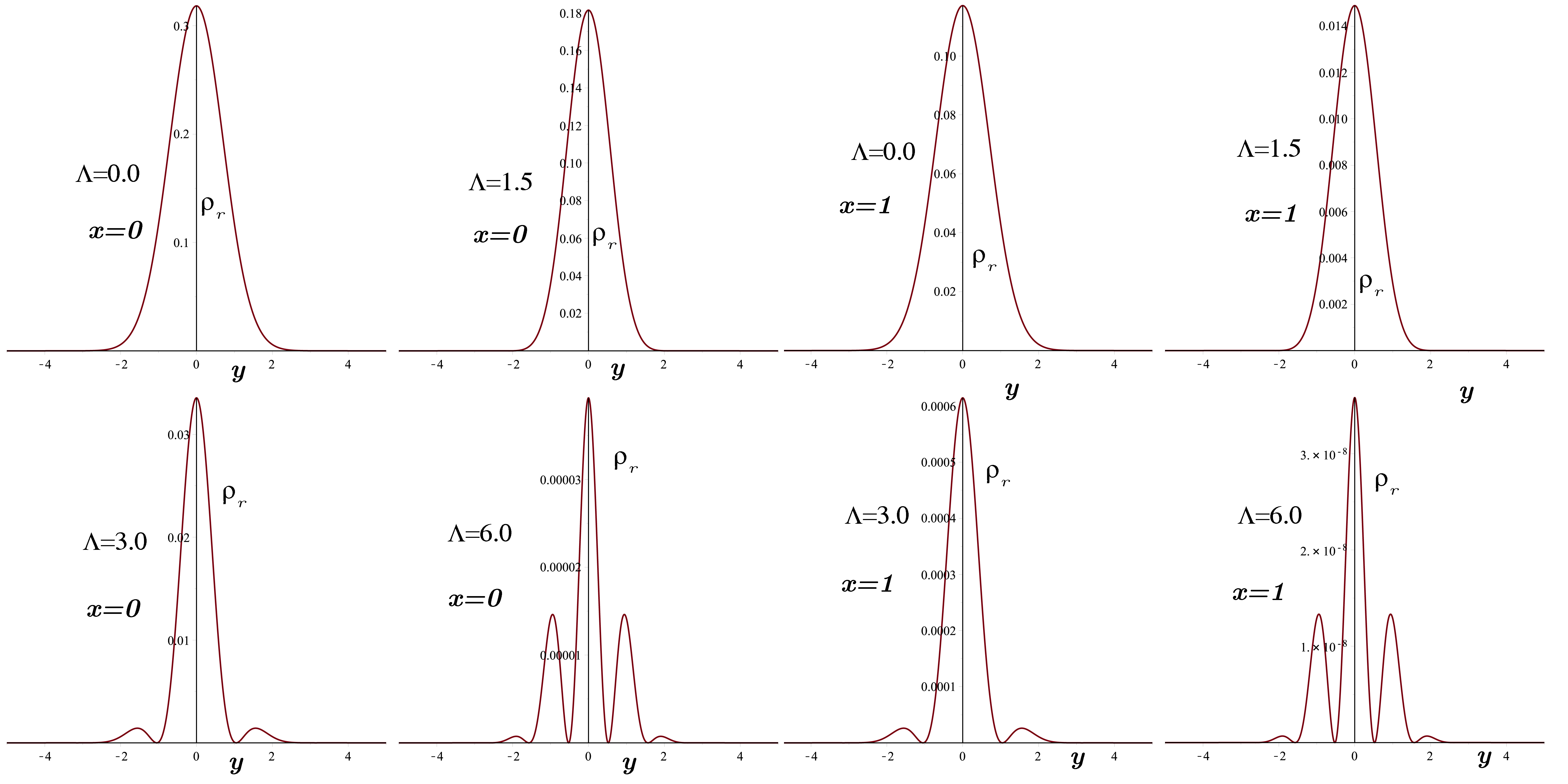} \caption{$\Re\left[\left(\psi_{00}\right)^{2}\right]$ in terms of $y$ for
different values of $\Lambda$ and two discrete values of $x=0$ and
$x=1$. }
\end{figure}

\begin{figure}
\includegraphics[scale=0.25]{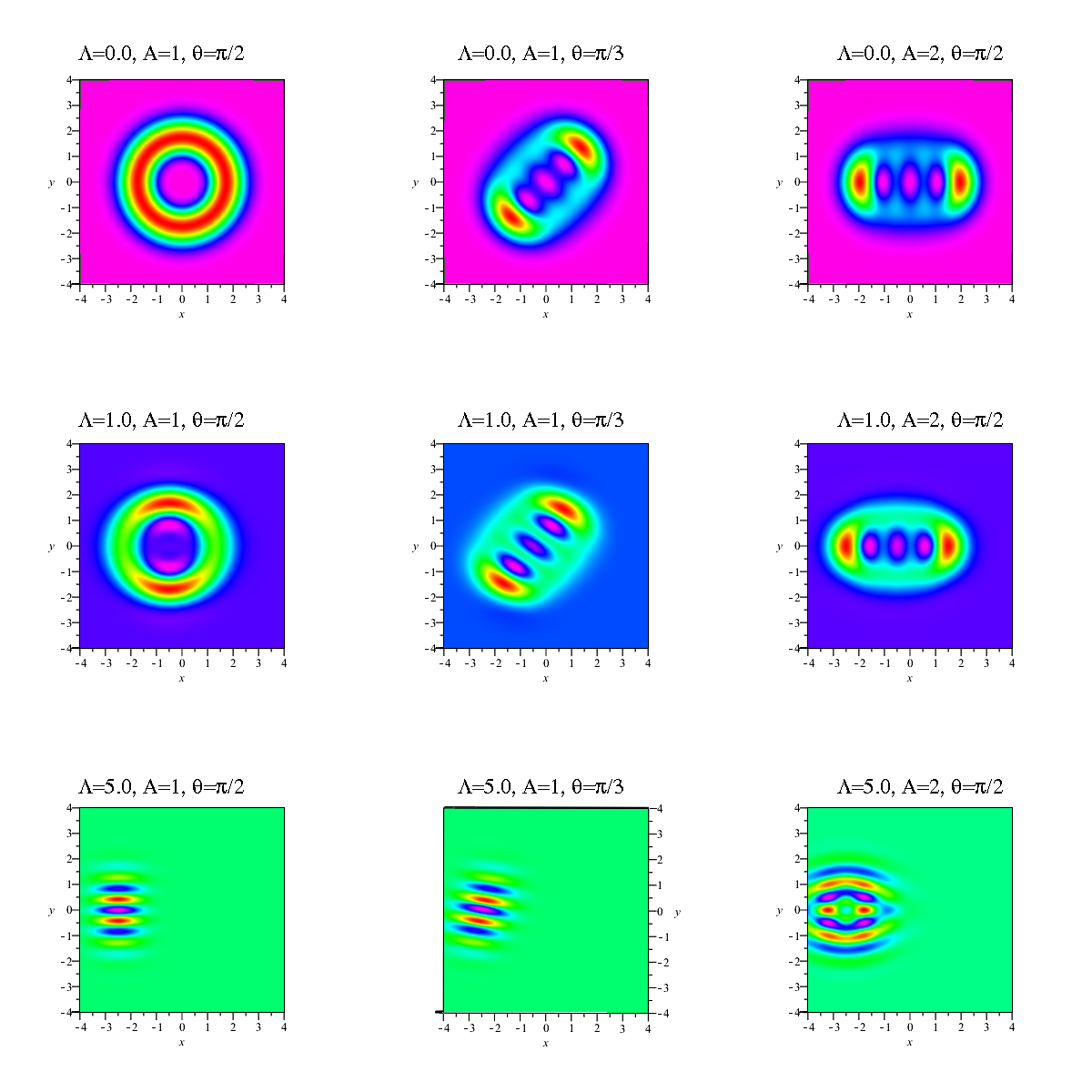} \caption{Plots of the $\Re\left(\Phi_{3}\right)^{2}$ in terms of $x$ and
$y$ with various value of $\Lambda,$ $A$ and $\theta,$ in accordance
with Eq. (33) with $N=3.$ The values of the parameters are specified
on each individual graphs.}
\end{figure}

\begin{figure}
\includegraphics[scale=0.25]{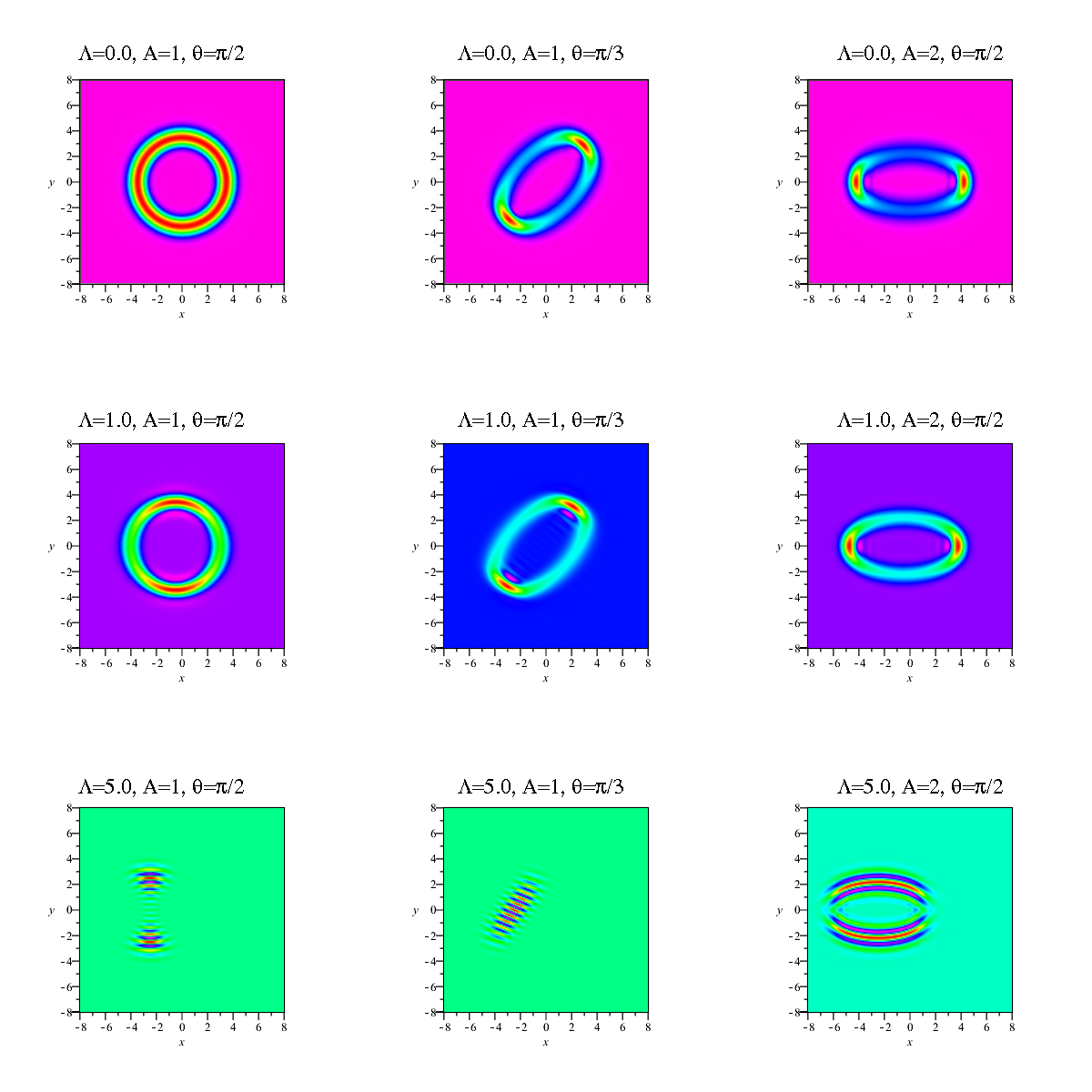} \caption{Plots of the $\Re\left(\Phi_{12}\right)^{2}$ in terms of $x$ and
$y$ with various value of $\Lambda,$ $A$ and $\theta,$ in accordance
with Eq. (29) with $N=12.$ The values of the parameters are specified
on the each graph.}
\end{figure}

\section{Results}

The $\Pi\mathcal{T}$-symmetric Hamiltonian with a complexified 2D
harmonic oscillator is considered within the time independent Schrödinger
equation. Wavefunctions ascertain in complex pattern due to the complex
argument of the Hermite polynomial for the $y$ component. Normalization
of the outcome wavefunctions are carried out based on the determination
of the $\mathcal{C}\Pi\mathcal{T}$-symmetry which is a weaker constraint
in comparison to the Hermitian Hamiltonian. Based on the definition
of the $\mathcal{C}\Pi\mathcal{T}$-operator and the explicit form
of the $\mathcal{C}\Pi\mathcal{T}$-normalized eigenfunctions given
by 
\begin{equation}
\psi_{nm}\left(x,y\right)=\dfrac{e^{-\left(x^{2}+y^{2}+\Lambda\left(x+iy\right)\right)/2}}{\sqrt{2^{n}2^{m}n!m!\pi}}H_{n}\left(x+\frac{\Lambda}{2}\right)H_{m}\left(y+i\frac{\Lambda}{2}\right)
\end{equation}
with $n,m=\left\{ 0,1,2,..,12\right\} $, one can easily show that
\begin{equation}
\int_{-\infty}^{\infty}\int_{-\infty}^{\infty}\left(\mathcal{C}\Pi\mathcal{T}\psi_{nm}\left(x,y\right)\right)\psi_{nm}\left(x,y\right)dxdy=\int_{-\infty}^{\infty}\int_{-\infty}^{\infty}\left(\psi_{nm}\left(x,y\right)\right)^{2}dxdy=1.
\end{equation}
We note that $\left(\psi_{nm}\left(x,y\right)\right)^{2}=\left\vert \psi_{nm}\left(x,y\right)\right\vert ^{2}$
for $\Lambda=0$ - which is nothing but the standard probability density
of the real 2D harmonic oscillator - while for $\Lambda\neq0,$ $\left(\psi_{nm}\left(x,y\right)\right)^{2}\neq\left\vert \psi_{nm}\left(x,y\right)\right\vert ^{2}$.
These suggest that we assume the probability density for the general
case to be of the form of $\left(\psi_{nm}\left(x,y\right)\right)^{2}$
other than $\left\vert \psi_{nm}\left(x,y\right)\right\vert ^{2}.$
Furthermore, the normalization condition (37) implies that 
\begin{equation}
\int_{-\infty}^{\infty}\int_{-\infty}^{\infty}\Re\left[\left(\psi_{nm}\left(x,y\right)\right)^{2}\right]dxdy=1
\end{equation}
while
\begin{equation}
\int_{-\infty}^{\infty}\int_{-\infty}^{\infty}\Im\left[\left(\psi_{nm}\left(x,y\right)\right)^{2}\right]dxdy=0
\end{equation}
which indicates that $\ Re\left[\left(\psi_{nm}\left(x,y\right)\right)^{2}\right]$
carries information about the particle in the real space. Hence, in
order to investigate the influence of the parameter $\Lambda$, one
may plot $\Re\left[\left(\psi_{nm}\left(x,y\right)\right)^{2}\right]$
with different values of $\Lambda$ and the results may be compared
with the actual probability density corresponding to $\Lambda=0$.
This is what we will do in the sequel. \\

Plots of $\Re\left[\left(\psi_{nm}\left(x,y\right)\right)^{2}\right]$
in terms of $x$ and $y$ for different values of $\Lambda$ are displayed
in Fig. 2. The first row illustrates the probability density considering
$\Lambda=0$. The lower ones depict the deformation due to imposing
the complexified potential into the system such that $\Lambda$ determines
the strength of the complexity. Fig. 3 depicts the real part of the
probability density of the ground state of the complexified harmonic
oscillator in terms of $y$ with different values of $\Lambda$ at
$x=0$ and $x=1$. While the effect of $\Lambda$ is easily seen in
this figure, the influence of $x$ should be found in the numerical
values on the graphs. Furthermore, by superposing 12 eigenstates of
the $\Pi\mathcal{T}$-symmetric Hamiltonian introduced earlier, we
found the stationary coherent state of the system. Similarly, as it
was mentioned for the normalization of the wavefunction $\psi_{nm}\left(x,y\right)$,
$\mathcal{C}\Pi\mathcal{T}$-operator is employed to normalize the
corresponding coherent states. In this respect, we assume that the
probability density of $\Phi_{N}$ is given by $\Re\left[\left(\Phi_{N}\right)^{2}\right]$.
The results are shown in Fig. 4 and Fig. 5 where we plot $\Re\left[\left(\Phi_{N}\right)^{2}\right]$
in terms of $x$ and $y$ for $N=3$ and $N=12$, respectively. It
is observed that variation of the amplitude and phase difference change
the form of the elliptical behavior.

\section{Conclusion}

In this paper, we studied the 2D-non-Hermitian Hamiltonian with the
complex potential given in Eq. (7). We found the eigenvalues and eigenfunctions
of the Hamiltonian analytically. Since the corresponding Hamiltonian
is $\Pi\mathcal{T}$-symmetric, the energy spectrum is real while
the energy eigenfunctions are complex. To resolve the negative norm
of the eigenfunctions, we employed the concepts of the charge operator
$\mathcal{C}$ and the so called $\mathcal{C}\Pi\mathcal{T}$-norm.
Furthermore, we carried on this work to find the coherent states of
the complex wave functions of the complexified 2D-harmonic oscillator.
The time-independent part of the coherent state describes the deformed-elliptical
motion of the wave packet without spreading behavior. Finally we plot
$\Re\left[\left(\psi_{nm}\left(x,y\right)\right)^{2}\right]$ and
$\Re\left[\left(\Phi_{N}\right)^{2}\right]$ with different configurations
to observe the effect of the non-Hermiticity parameter $\Lambda.$
Since with $\Lambda=0,$ $\Re\left[\left(\psi_{nm}\left(x,y\right)\right)^{2}\right]$
and $\Re\left[\left(\Phi_{N}\right)^{2}\right]$ reduce to the standard
probability densities i.e., $\left\vert \psi_{nm}\left(x,y\right)\right\vert ^{2}$
and $\left\vert \Phi_{N}\right\vert ^{2},$ respectively, and also
they satisfy (38) we have considered them to be the reasonable candidates
for the probability densities in the real $xy$-plane. Figs. 4 and
5 reveal the effects of the parameter $\Lambda$ in the probability
density of the coherent state of the complexified 2D harmonic oscillator
in the form of either diffusions or dispersions.

\bigskip{}

\end{document}